\documentclass[aps,reprint,nofootinbib,amsmath,amssymb,floatfix]{revtex4-2}
\usepackage[T1]{fontenc}
\usepackage{graphicx,bm,booktabs,xcolor}
\usepackage[colorlinks=true,linkcolor=blue,urlcolor=blue,citecolor=blue]{hyperref}
\newcommand{\avg}[1]{\langle #1\rangle}

\newcommand{\eu}{\hat{\bm e}}
\newcommand{\dd}{\mathrm d}
\newcommand{\ii}{\mathrm i}
\begin{document}
\title{Cosmological Constraints on Dark Photon Dark Matter \\ from Sustained Resonant Heating}
\author{Lu-Yao Li}
\author{Jing Lin}
\author{Chuan-Yang Xing}
\email{cyxing@upc.edu.cn}
\author{Ying-Jie Zhang}
\affiliation{College of Science, China University of Petroleum (East China), Qingdao 266580, China}
\begin{abstract}
Dark photon dark matter can resonantly excite plasma waves during cosmic evolution. Once these waves reach sufficiently large amplitudes, nonlinear interactions transfer energy from the spatially uniform mode to finite-wavelength sidebands, suppressing the initial resonant growth. Using driven Zakharov simulations and a model of wave-packet collapse, we show that the residual uniform mode nevertheless continues to absorb power from the dark field. This power feeds the sidebands, which nonlinear focusing concentrates into localized wave packets that collapse and dissipate their energy as heat. The resulting quasi-steady cycle of energy injection and dissipation sustains plasma heating well beyond saturation. Adopting an intergalactic heating budget of $1\,{\rm eV}$ per baryon over redshifts $2\le z\le 6$ yields an approximate upper bound of $\epsilon \lesssim 2.6\times10^{-12}$ on the kinetic mixing parameter for dark photon masses of $(0.9\text{--}3.2)\times10^{-13}\,{\rm eV}$.
\end{abstract}
\maketitle
\section{Introduction}
Unraveling the microscopic nature of dark matter (DM) requires identifying how it couples to Standard Model fields. A massive dark photon is a minimal and well-motivated candidate for cold DM, interacting with electromagnetic currents via kinetic mixing with the ordinary photon~\cite{Arias:2012az}. In the ultralight regime, it behaves as a coherently oscillating classical field that acts on charged particles through a weak effective electric field. Because collective plasma oscillations can strongly amplify the response to this weak driving, plasmas across diverse astrophysical and terrestrial environments serve as sensitive probes of dark photon DM~\cite{An:2020jmf,An:2023wij,An:2024wmc,Hardy:2022ufh,Beadle:2024jlr,Zatini:2025pob}. On cosmological scales, the expanding intergalactic plasma is especially sensitive because its thermal history records energy transferred from the dark field throughout cosmic evolution.

A primary mechanism for this energy transfer is resonant conversion. As the cosmic electron density evolves, the plasma frequency can pass through the dark photon oscillation frequency, resonantly exciting Langmuir waves that subsequently damp and heat the gas. Because DM stores roughly $5\,{\rm GeV}$ of energy per baryon whereas depositing just $1\,{\rm eV}$ per baryon noticeably alters the temperature of the intergalactic medium, even a tiny conversion fraction yields stringent cosmological bounds on the kinetic mixing parameter $\epsilon$~\cite{Arias:2012az,McDermott:2019lch}. Accounting for spatial density fluctuations refines when and where these resonant crossings occur~\cite{Caputo:2020bdy,Witte:2020rvb,Caputo:2020rnx}. Furthermore, hydrodynamical simulations have linked the resulting energy deposition to Lyman-$\alpha$ forest observations, both as a possible explanation for excess low-redshift heating and as a means of constraining $\epsilon$~\cite{Bolton:2022hpt,Trost:2024ciu}.

Recently, however, Hook, Huang, and Shalaby argued that nonlinear plasma dynamics invalidates these cosmological bounds~\cite{Hook:2025pbn}. Standard conversion calculations assume a linear plasma response on a fixed background density, but growing Langmuir waves exert ponderomotive forces that modify the local density and shift the local plasma frequency away from resonance with the dark field. Using 1D kinetic simulations, they showed that these self-generated density fluctuations prematurely saturate resonant wave growth and concluded that the total energy transferred to the plasma is far too small to affect the thermal history of the Universe.

Reassessing these constraints requires tracking the plasma beyond the initial suppression of resonant growth, because cosmological heating depends on the energy deposited throughout the entire resonance crossing, not merely before saturation. In the 1D collisionless simulations of Ref.~\cite{Hook:2025pbn}, post-saturation energy injection from the dark field remains weak. In 2D and 3D, by contrast, Langmuir wave packets can undergo collapse, transferring wave energy to short scales where it dissipates as heat~\cite{RN3,Newman}. In a quasi-steady regime, such dissipation is balanced by continued energy injection from the dark field, thereby sustaining plasma heating after saturation.

In this work, we demonstrate that sustained resonant heating revives cosmological constraints on dark photon DM. Based on the Zakharov equations, we find that after the initial coherent growth saturates, the plasma settles into a quasi-steady turbulent state. In this regime, a residual coherent component of the uniform Langmuir mode continues to draw power from the dark field, balancing the dissipation of finite-wavelength modes. Combining this response with a model of Langmuir-wave collapse, we estimate the heat deposited during a cosmological resonance crossing. Requiring this heating not to exceed $1\,{\rm eV}$ per baryon in the intergalactic medium at redshifts $2\le z\le 6$, we obtain an upper bound of $\epsilon \lesssim 2.6\times 10^{-12}$ for dark photon masses in the range $(0.9\text{--}3.2)\times 10^{-13}\,{\rm eV}$. Although this bound is weaker than linear-response estimates, our results establish that nonlinear plasma turbulence sustains sufficient energy transfer to preserve cosmological sensitivity to dark photon DM.

We use natural units with $\hbar=c=k_B=1$.

\section{Nonlinear Evolution of Resonantly Driven Plasma Waves}
Through kinetic mixing, the dark photon drives charged particles in the plasma via a nearly uniform effective electric field~\cite{Horns:2012jf,Xing:2026yvf,Li:2026mqj},
\begin{equation}
 \bm E_{\rm DP}(t)=\operatorname{Re}\!\left[\bm{\mathcal E}_{\rm DP}e^{-\ii\omega_{A'}t}\right],\quad
 |\bm{\mathcal E}_{\rm DP}|=\epsilon\sqrt{2\rho_{\rm DP}}.
 \label{eq:physical-drive}
\end{equation}
Here $t$ is the physical time, $\bm{\mathcal E}_{\rm DP}$ is the complex field amplitude, $\epsilon$ is the kinetic mixing parameter, $\rho_{\rm DP}$ is the dark photon energy density, and $\omega_{A'} \approx m_{A'}$ is the oscillation frequency, with $m_{A'}$ denoting the dark photon mass. In a plasma composed of electrons and singly charged ions with background electron density $n_0$, the electron plasma frequency is
\begin{equation}
 \omega_p=\sqrt{\frac{n_0e^2}{m_e}},
 \label{eq:plasma-frequency}
\end{equation}
where $e$ is the magnitude of the electron charge and $m_e$ is the electron mass. When $\omega_{A'}$ matches $\omega_p$, the oscillating dark photon field can resonantly excite the spatially uniform Langmuir mode in the plasma.

To capture the nonlinear feedback of these waves on the plasma, we employ the Zakharov framework, which couples fast Langmuir oscillations to slow density variations~\cite{Zak,Rob}. We write the total effective electric field as $\bm E_{\rm tot}(\bm x,t)=\operatorname{Re}\!\left[\bm{\mathcal E}(\bm x,t)e^{-\ii\omega_pt}\right]$, where $\bm x$ is the physical position and $\bm{\mathcal E}(\bm x,t)$ is the slowly varying envelope of the combined drive and self-consistent plasma response. At resonance, the envelope couples to a slow, approximately quasineutral density perturbation $\delta n$ through the Zakharov equations with an external dark photon drive,
\begin{subequations}
\begin{gather}
 \ii\partial_t\bm{\mathcal E}
 =-\frac{3T_e}{2m_e\omega_p}\nabla^2\bm{\mathcal E}
 +\frac{\omega_p}{2n_0}\mathcal P_{\rm L}(\delta n\,\bm{\mathcal E})
 +\frac{\omega_p}{2}\bm{\mathcal E}_{\rm DP}
 -\ii\widehat\Gamma\bm{\mathcal E},
 \label{eq:physical-field}\\
 \left(\partial_t^2+2\widehat\Gamma_{\rm s}\partial_t
 -c_s^2\nabla^2\right)\delta n
 =\frac{1}{4m_i}\nabla^2|\bm{\mathcal E}|^2.
 \label{eq:physical-density}
\end{gather}
\end{subequations}
Here $T_e$ and $T_i$ are the electron and ion temperatures, $m_i$ is the ion mass, and $c_s$ is the ion-acoustic sound speed with $c_s^2=(T_e+3T_i)/m_i$. The damping operators $\widehat\Gamma$ and $\widehat\Gamma_{\rm s}$ act on Langmuir and ion-acoustic waves, respectively, while $\mathcal P_{\rm L}$ projects onto the longitudinal component and preserves the uniform mode. Physically, the ponderomotive pressure on the right-hand side of Eq.~\eqref{eq:physical-density} expels plasma from regions of high wave intensity, producing a density perturbation that shifts the local plasma frequency by
\begin{equation}
 \delta\omega_p\simeq\frac{\omega_p}{2}\frac{\delta n}{n_0}.
 \label{eq:density-detuning}
\end{equation}
Through the $\delta n\,\bm{\mathcal E}$ term in Eq.~\eqref{eq:physical-field}, this frequency shift feeds back on the Langmuir waves, disrupting their coherent resonant growth and allowing wave energy to concentrate in density depressions.

\begin{figure}[!t]
 \includegraphics[width=\columnwidth]{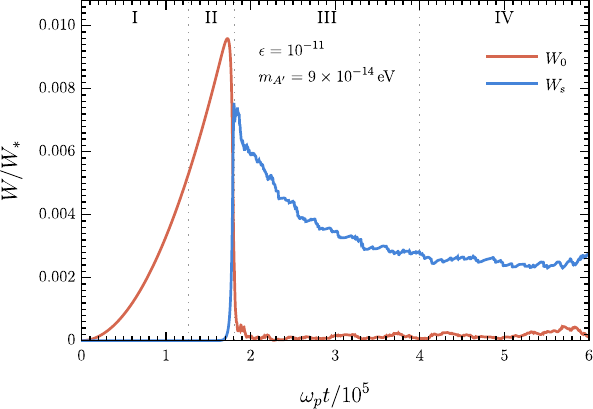}
 \caption{\label{fig:evolution}Langmuir energy densities of the uniform mode ($W_0$) and the sidebands ($W_{\rm s}$) in units of $W_*=n_0T_e/2$, from a 2D driven Zakharov simulation with $\epsilon=10^{-11}$ and $m_{A'}=9\times10^{-14}\,{\rm eV}$. Stages~I--IV denote coherent growth, modulational instability, sideband decay, and the sustained driven state.}
\end{figure}

Because the dark photon directly drives only the uniform mode while nonlinear coupling transfers energy to finite wavelengths, it is useful to separate the field envelope into its uniform component and finite-wavelength sidebands. With $\avg{\cdot}$ denoting a spatial average, these envelopes are
\begin{equation}
 \bm{\mathcal E}_0=\avg{\bm{\mathcal E}},\qquad
 \bm{\mathcal E}_{\rm s}=\bm{\mathcal E}-\bm{\mathcal E}_0,
 \label{eq:physical-modes}
\end{equation}
where the subscripts $0$ and ${\rm s}$ label the uniform mode and sidebands, respectively. Once the resonantly enhanced plasma field dominates the directly applied field, their leading-order Langmuir energy densities are
\begin{equation}
 W_0\simeq\frac{1}{2}|\bm{\mathcal E}_0|^2,
 \qquad W_{\rm s}\simeq\frac{1}{2}
 \avg{|\bm{\mathcal E}_{\rm s}|^2}.
 \label{eq:physical-mode-energies}
\end{equation}
These expressions include both the electric field energy and the equal kinetic contribution from electron oscillations, and therefore represent energy stored in the waves rather than heat already deposited in the plasma. Tracking $W_0$ and $W_{\rm s}$ in a 2D driven Zakharov simulation reveals four distinct dynamical stages, as shown in Fig.~\ref{fig:evolution}.

\textbf{I. Coherent resonant growth.}
Initially, the response is dominated by the uniform mode, while the sidebands remain weak. Neglecting damping of the uniform mode and feedback from density fluctuations, Eq.~\eqref{eq:physical-field} reduces to
\begin{equation}
 \ii\frac{\dd\bm{\mathcal E}_0}{\dd t}
 =\frac{\omega_p}{2}\bm{\mathcal E}_{\rm DP}.
 \label{eq:physical-uniform-drive}
\end{equation}
Integrating Eq.~\eqref{eq:physical-uniform-drive} from an initial time $t_0$ yields
\begin{equation}
 \bm{\mathcal E}_0(t)=\bm{\mathcal E}_0(t_0)
 -\frac{\ii\omega_p}{2}\bm{\mathcal E}_{\rm DP}(t-t_0).
 \label{eq:physical-linear-growth}
\end{equation}
The resonantly generated field therefore grows linearly with time. Once this contribution dominates the initial field, the uniform-mode energy density grows as $W_0\propto(t-t_0)^2$, accounting for the early rise in Fig.~\ref{fig:evolution}.

\textbf{II. Modulational instability and sideband growth.}
As the uniform oscillation strengthens, its coupling to density fluctuations becomes important. A density perturbation scatters the uniform field into finite-wavelength waves. Their interference with the uniform field produces a spatial modulation of the wave intensity, whose ponderomotive pressure further amplifies the density perturbation. This feedback drives the modulational instability and the rapid growth of the sidebands~\cite{Zak}. Simultaneously, the growing density perturbations shift the local plasma frequency, disrupting the coherent response to the dark field~\cite{Hook:2025pbn}. In Fig.~\ref{fig:evolution}, the uniform mode reaches a maximum and begins to decline as the sidebands become dominant. The initial resonant growth is thus suppressed, while a substantial fraction of the wave energy is redistributed into finite-wavelength modes.

\textbf{III. Sideband decay and wave-packet dissipation.}
After reaching their maximum, the sidebands lose much of the energy accumulated during the instability. In 2D and 3D, the same ponderomotive feedback continues to concentrate the sideband waves trapped in density depressions. Once a localized wave packet is sufficiently intense, self-focusing overcomes thermal dispersion and drives runaway contraction toward progressively shorter spatial scales, a process known as Langmuir wave collapse~\cite{RN3,Newman}. As the packet shrinks to sufficiently small scales, thermal electrons traverse its coherent field within a plasma oscillation period and extract energy from the wave. This transit-time damping heats the plasma while depleting the packet and arresting collapse~\cite{Rob} (see Appendix~\ref{app:kinetic-packet} for a particle-in-cell simulation~\cite{WarpX} of wave-packet collapse and arrest). Within the fluid Zakharov description, we follow Robinson and Newman~\cite{RN} and model this short-scale kinetic absorption by the damping rate
\begin{equation}
 \Gamma(k)\simeq\Gamma_c\lambda_{\rm De}^2(k^2-k_c^2)\Theta(k-k_c),
 \label{eq:physical-damping}
\end{equation}
where $k$ is the physical wave number and $\lambda_{\rm De}=\sqrt{T_e/m_e}/\omega_p$ is the Debye length. We use the cutoff $k_c=0.25/\lambda_{\rm De}$ and coefficient $\Gamma_c=4\omega_p$ from Ref.~\cite{RN}. This enhanced short-scale damping captures the rapid dissipation of collapsing wave packets, a channel that is absent in 1D~\cite{Rob}. In Fig.~\ref{fig:evolution}, the stepwise decline of $W_{\rm s}$ during Stage~III reflects discrete collapse-and-arrest events of individual wave packets.

\textbf{IV. Sustained driven state.}
However, this initial decay does not exhaust the sideband population. At late times, the plasma settles into a quasi-steady turbulent state dominated by finite-wavelength sidebands alongside a smaller uniform component. Although the stored wave energy remains bounded, plasma heating continues: the dark field drives the residual uniform mode, density fluctuations scatter this energy into the sidebands, and collapsing wave packets repeatedly dissipate it as heat. Quantifying this sustained heating rate requires a power-balance analysis as in the next section.

\section{Energy Injection and Dissipation}
\begin{figure}[!t]
 \includegraphics[width=\columnwidth]{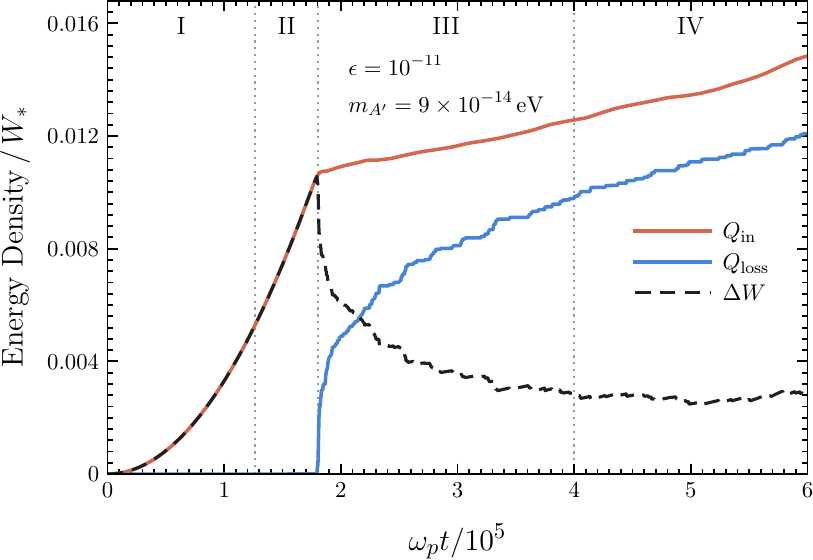}
 \caption{\label{fig:budget}Cumulative injected energy density $Q_{\rm in}$, cumulative dissipated energy density $Q_{\rm loss}$, and change in stored wave energy density $\Delta W$, all in units of $W_*=n_0T_e/2$. The dark photon parameters and stage boundaries are the same as in Fig.~\ref{fig:evolution}.}
\end{figure}

The dark photon drive supplies power density $\mathcal P_{\rm in}$ to the plasma. As shown in Fig.~\ref{fig:budget}, the cumulative input $Q_{\rm in}(t)=\int_0^t\mathcal P_{\rm in}(t')\,\dd t'$ grows quadratically with time during Stages~I and~II, as expected for coherent resonant driving~\cite{Hook:2025pbn}, and transitions to an approximately linear rise in Stage~IV. In this late-time regime, the dark photon drive feeds the uniform mode at rate $\mathcal P_{\rm in}$, density scattering transfers wave energy from the uniform mode to the sidebands at rate $\mathcal P_{\rm tr}$, and short-scale damping removes sideband energy at rate $\mathcal P_{\rm loss}$. Because the stored energy in each component is nearly stationary on average, their separate energy budgets give
\begin{equation}
 \overline{\mathcal P}_{\rm in}\simeq
 \overline{\mathcal P}_{\rm tr}\simeq
 \overline{\mathcal P}_{\rm loss},
 \label{eq:balance}
\end{equation}
where overbars denote late-time averages. This balance between injection and dissipation is directly visible in Fig.~\ref{fig:budget}: in Stage~IV, the cumulative dissipation $Q_{\rm loss}(t)=\int_0^t\mathcal P_{\rm loss}(t')\,\dd t'$ rises with nearly the same slope as $Q_{\rm in}$, while the change in stored wave energy density $\Delta W$ fluctuates about a steady level.

To obtain a theoretical estimate of the mean power densities in Eq.~\eqref{eq:balance}, we draw on the two-component model of strong Langmuir turbulence~\cite{Rob,RN}. In this model, each of the three mean power densities is expressed in terms of the mean energy densities of the uniform and sideband components, $\overline W_0$ and $\overline W_{\rm s}$, respectively. Imposing the power balance in Eq.~\eqref{eq:balance} then determines $\overline W_0$, $\overline W_{\rm s}$, and the scaling of the steady-state injection power with $\epsilon$.

From the drive term in Eq.~\eqref{eq:physical-uniform-drive} with real $\bm{\mathcal E}_{\rm DP}$, the mean injection power density is
\begin{equation}
 \overline{\mathcal P}_{\rm in}=
 \frac{\epsilon\sqrt{2\rho_{\rm DP}}}{2}\,
  \left|\operatorname{Im}\overline{\bm{\mathcal E}_0}\right|\,\omega_p ,
 \label{eq:steady-injection}
\end{equation}
where $\overline{\bm{\mathcal E}_0}$ is the mean uniform field, taken to be parallel to the drive polarization. When fluctuations of the uniform mode are negligible, Eq.~\eqref{eq:physical-mode-energies} gives $\left|\overline{\bm{\mathcal E}_0}\right|\simeq\sqrt{2\overline W_0}$. We therefore write $\left|\operatorname{Im}\overline{\bm{\mathcal E}_0}\right|\simeq C_{\rm in}\sqrt{2\overline W_0}$, where the dimensionless factor $C_{\rm in}$ accounts for the phase of $\overline{\bm{\mathcal E}_0}$ relative to the drive. For the parameters used here, the model in Appendix~\ref{app:density-scattering} gives $C_{\rm in}\simeq0.505$ in 2D and $C_{\rm in}\simeq0.590$ in 3D.

Energy from the driven uniform mode is scattered into the sidebands by density fluctuations via the $\delta n\,\bm{\mathcal E}$ coupling in Eq.~\eqref{eq:physical-field}. A larger uniform-mode energy density $\overline W_0$ supplies more energy to be scattered, thereby increasing the transfer power density. Meanwhile, a larger sideband energy density $\overline W_{\rm s}$ enhances the scattering rate itself, as the sideband wave packets drive stronger density fluctuations. We therefore estimate the mean transfer power density as
\begin{equation}
\overline{\mathcal P}_{\rm tr} \simeq C_{\rm tr}\,\overline W_0\,\overline W_{\rm s}\,\frac{\omega_p}{W_*},
\label{eq:steady-transfer}
\end{equation}
where $W_*=n_0T_e/2$ is the reference energy density and $C_{\rm tr}$ is a dimensionless transfer coefficient determined by the spatial spectrum and correlation time of the density fluctuations. For the present parameters, we obtain $C_{\rm tr}\simeq 0.0366$ in 2D and $C_{\rm tr}\simeq 0.0347$ in 3D, as derived in Appendix~\ref{app:density-scattering}.

Energy transferred to the sidebands is dissipated when short-scale damping arrests the collapse of Langmuir wave packets~\cite{Newman}. In the nucleation and collapse theory of Refs.~\cite{RN,Rob}, the mean dissipated power density is $\overline{\mathcal P}_{\rm loss}=(N_P/T_N)U_{\rm loss}$, where $N_P$ is the wave-packet number density, $T_N$ is the mean interval between collapses at a given site, and $U_{\rm loss}$ is the mean energy dissipated per collapse. As the total wave energy density $W=\overline W_0+\overline W_{\rm s}$ increases, packets begin to collapse at smaller widths, so their number density grows as $N_P\propto W^{D/2}$ (with $D$ the number of spatial dimensions) and the interval between collapses shortens as $T_N\propto W^{-1}$, while $U_{\rm loss}$ remains approximately independent of $W$. These scalings give~\cite{RN,Rob}
\begin{equation}
 \overline{\mathcal P}_{\rm loss}
 \simeq \alpha_D(\overline W_0+\overline W_{\rm s})^{1+D/2}
 \frac{\omega_p}{W_*^{D/2}},
 \label{eq:closure}
\end{equation}
where the dimensionless coefficients $\alpha_2\simeq0.002$ and $\alpha_3\simeq0.0063$ follow from rescaling the empirical values in Refs.~\cite{RN,Rob} to our plasma parameters.

\begin{figure}[!t]
 \includegraphics[width=\columnwidth]{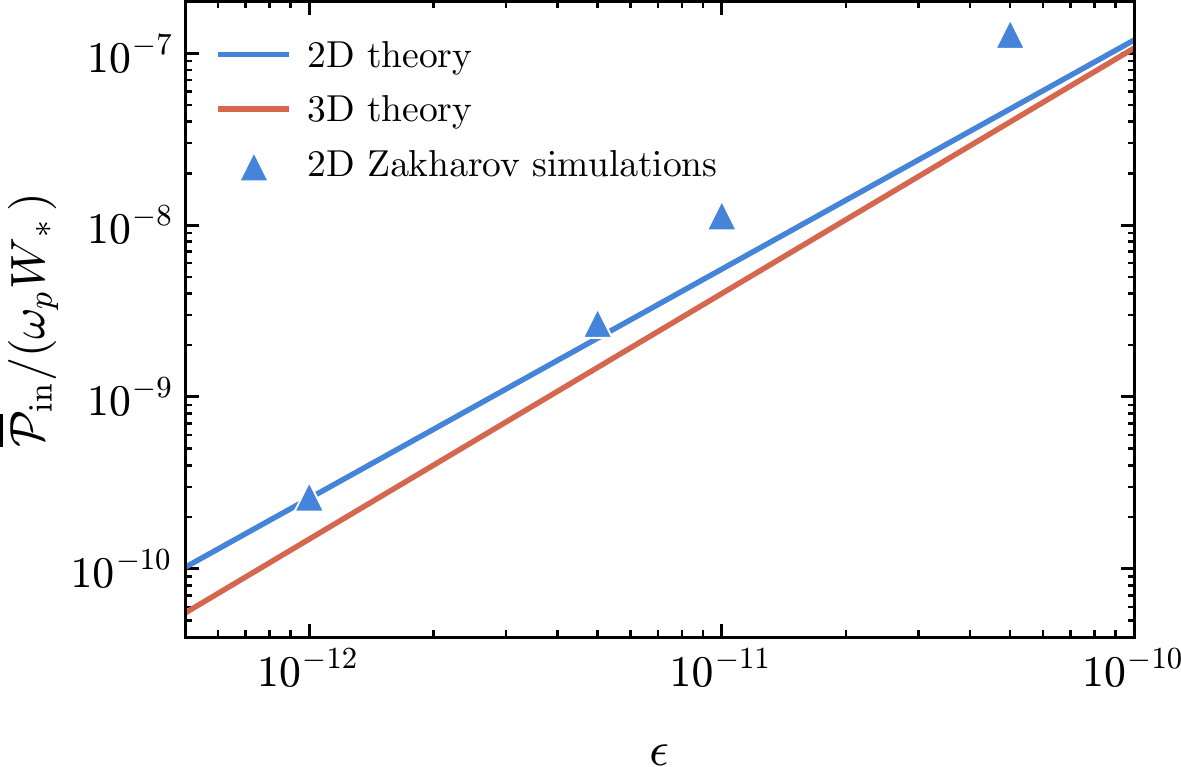}
 \caption{\label{fig:steady-power}Mean injection power density $\overline{\mathcal P}_{\rm in}$, in units of $\omega_p W_*$, as a function of the kinetic mixing parameter $\epsilon$. The curves show the predictions of Eq.~\eqref{eq:steady-power} in 2D and 3D, while the triangles denote the results of 2D driven Zakharov simulations.}
\end{figure}

In the quasi-steady state, balancing these three power densities determines $\overline W_0$ and $\overline W_{\rm s}$, as well as the dependence of the mean injection power density on $\epsilon$. For $\overline W_{\rm s}\gg\overline W_0$, this yields
\begin{equation}
 \overline{\mathcal P}_{\rm in}\simeq
 \begin{cases}
  0.46\,\epsilon^{4/3}\rho_{\rm DP}^{2/3}\omega_p W_*^{1/3}, & D=2,\\
  1.22\,\epsilon^{10/7}\rho_{\rm DP}^{5/7}\omega_p W_*^{2/7}, & D=3.
 \end{cases}
 \label{eq:steady-power}
\end{equation}
These two power laws are compared with our four 2D Zakharov simulations in Fig.~\ref{fig:steady-power}. The simulated mean injection power densities agree with the $D=2$ prediction up to an order-unity factor.

\begin{figure*}[!t]
 \includegraphics[width=0.95\textwidth]{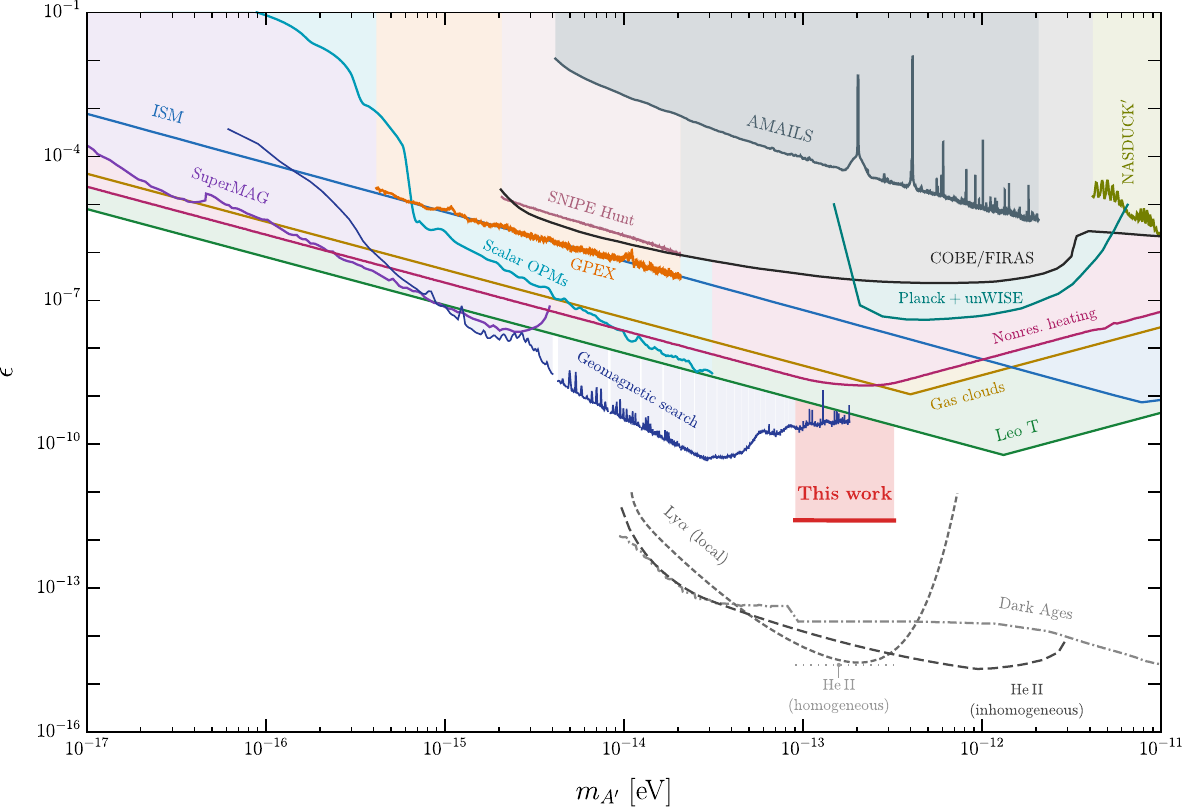}
 \caption{\label{fig:constraint}Upper limits on the kinetic mixing $\epsilon$ as a function of the dark photon mass $m_{A'}$. The red curve gives our heating bound for a heating budget of $1\,{\rm eV}$ per baryon over $2\leq z\leq6$. Gray curves show earlier resonant-heating bounds obtained without nonlinear effects~\cite{McDermott:2019lch,Caputo:2020bdy,Witte:2020rvb,Caputo:2020rnx}. Other colored curves indicate complementary astrophysical and direct-search limits~\cite{Dubovsky:2015cca,Wadekar:2019mpc,McDermott:2019lch,Caputo:2020bdy,Arsenadze:2024ywr,Chluba:2024wui,McCarthy:2024ozh,Fedderke:2021rrm,Friel:2024shg,Jiang:2023jhl,Sulai:2023zqw,Nomura:2025rfi,Zhao:2025kdw,Arza:2025kuh,Barir:2026awl}.}
\end{figure*}

\section{Cosmological Heating and Constraints}
As the Universe expands, the decreasing plasma frequency passes through the dark photon oscillation frequency at redshift $z_r$, where $\omega_p(z_r)=m_{A'}$, and the dark photon resonantly drives the plasma. The cumulative injected energy grows quadratically with time before saturation and linearly in the subsequent quasi-steady state. Since this quasi-steady phase lasts much longer than the initial growth stage, it dominates the total energy deposition, which can therefore be estimated by multiplying the 3D mean injection power density $\overline{\mathcal P}_{\rm in}$ in Eq.~\eqref{eq:steady-power} by the resonance duration. As $\omega_p$ continues to decrease, the plasma oscillation accumulates a phase difference relative to the drive. This phase difference reaches order unity on the timescale~\cite{Hook:2025pbn}
\begin{equation}
 t_{\rm res}\simeq[H(z_r)\,\omega_p(z_r)]^{-1/2},
 \label{eq:resonance-duration}
\end{equation}
where $H(z_r)$ is the Hubble rate at resonance~\cite{Planck:2018vyg}. Taking $t_{\rm res}$ as the duration of quasi-steady absorption, we obtain the heat deposited per baryon,
\begin{equation}
 Q_{\rm heat}\simeq\frac{\overline{\mathcal P}_{\rm in}\,t_{\rm res}}{n_b(z_r)},
 \label{eq:heatint}
\end{equation}
where $n_b(z_r)$ is the baryon number density at resonance.

The deposited heat raises the temperature of the intergalactic gas and thermally broadens its absorption lines~\cite{Schaye:1999vr,Bolton:2022hpt}. Consequently, the thermal history inferred from the Lyman-$\alpha$ forest places an upper bound on the energy injection given by Eq.~\eqref{eq:heatint}. Following McDermott and Witte~\cite{McDermott:2019lch}, we adopt a limit of $1\,{\rm eV}$ per baryon over $2\leq z\leq6$. This yields $\epsilon\lesssim2.6\times10^{-12}$ for a crossing at $z_r=2$, corresponding to a dark photon mass of $m_{A'}=9\times10^{-14}\,{\rm eV}$. This bound is nearly independent of $m_{A'}$ across the mass range $(0.9\text{--}3.2)\times10^{-13}\,{\rm eV}$, as shown by the red curve in Fig.~\ref{fig:constraint}.

Earlier calculations of resonant heating assumed a linear plasma response and neglected saturation effects~\cite{McDermott:2019lch,Caputo:2020bdy,Witte:2020rvb,Caputo:2020rnx}. In this regime, the cumulative energy input grows quadratically with time, as in Stage~I, throughout the full resonance duration $t_{\rm res}$, resulting in a much larger total energy deposition at a fixed $\epsilon$~\cite{Hook:2025pbn}. For the same heating budget, this quadratic energy injection during He~II reionization~\cite{McDermott:2019lch} therefore yields an upper bound roughly three orders of magnitude tighter than our quasi-steady estimate, as shown by the gray dotted curve in Fig.~\ref{fig:constraint}. The remaining gray curves represent extensions of this linear-response treatment to an inhomogeneous intergalactic medium~\cite{Caputo:2020bdy,Witte:2020rvb,Caputo:2020rnx} or other cosmic epochs~\cite{McDermott:2019lch}. 

For comparison, the other colored curves in Fig.~\ref{fig:constraint} show complementary bounds from nonresonant gas heating~\cite{Dubovsky:2015cca,Wadekar:2019mpc,McDermott:2019lch}, CMB photon conversion~\cite{Caputo:2020bdy,Arsenadze:2024ywr,Chluba:2024wui,McCarthy:2024ozh}, and direct searches~\cite{Fedderke:2021rrm,Friel:2024shg,Jiang:2023jhl,Sulai:2023zqw,Nomura:2025rfi,Zhao:2025kdw,Arza:2025kuh,Barir:2026awl}.

\section{Conclusions and Discussion}
We show that resonant heating of the cosmic plasma by dark photon dark matter persists beyond the initial saturation of coherent wave growth. Driven Zakharov simulations reveal a quasi-steady turbulent state in which the residual uniform mode continues to absorb power from the dark photon, balancing transfer to finite-wavelength sidebands and dissipation through wave-packet collapse. Combining this power balance with a collapse dissipation model yields an approximate upper bound of $\epsilon\lesssim 2.6\times10^{-12}$ for $m_{A'}\in(0.9\text{--}3.2)\times10^{-13}\,{\rm eV}$.

Direct 3D simulations incorporating a time-varying plasma frequency will refine the predicted heating across resonance crossings. Extending this nonlinear heating framework to an inhomogeneous intergalactic medium and other cosmic epochs will enable a broader reassessment of cosmological bounds across a wider range of dark photon masses.

\begin{acknowledgments}
This work was supported by the National Natural Science Foundation of China (Grant No.~12605182), the Fundamental Research Funds for the Central Universities (No.~24CX06048A), and the Taishan Scholar Young Talent Program (Grant No.~tsqn202507117).
\end{acknowledgments}

\bibliography{references}
\onecolumngrid
\clearpage
\twocolumngrid
\appendix
\renewcommand{\theHequation}{\theequation}
\section{Kinetic Simulation of Wave-Packet Collapse and Dissipation}\label{app:kinetic-packet}
\setcounter{equation}{0}\renewcommand{\theequation}{A\arabic{equation}}
To illustrate the collapse, arrest, and kinetic dissipation of Langmuir wave packets, we perform a 2D collisionless particle-in-cell simulation with WarpX~\cite{WarpX}. The plasma is initialized with $T_e=T_i=10^4\,{\rm K}$, the physical proton-to-electron mass ratio $m_i/m_e=1836.15$, and a spatially uniform electric field giving an electron quiver speed of $0.5\sqrt{T_e/m_e}$. The periodic square domain has side length $192.7\lambda_{\rm De}$ on a $576^2$ grid, with $256$ particles per species per cell and time step $\omega_p\Delta t=0.025$.

\begin{figure}[!t]
 \centering
 \includegraphics[width=.95\columnwidth]{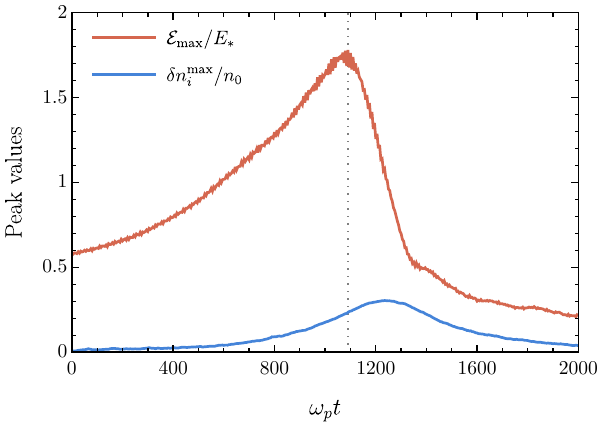}
 \caption{\label{fig:warpx}Evolution of the peak electric-field envelope $\mathcal E_{\max}/E_*$ (with $E_*=\sqrt{n_0T_e}$) and deepest ion-density depression $\delta n_i^{\max}/n_0$ of a collapsing Langmuir wave packet in the 2D WarpX simulation. The vertical dotted line marks the field maximum at $\omega_p t=1090$, signaling the onset of collapse arrest.}
\end{figure}

As in the Zakharov evolution, the initial uniform oscillation breaks up via the modulational instability into localized wave packets trapped in ion density depressions. Fig.~\ref{fig:warpx} follows the peak electric field envelope $\mathcal E_{\max}/E_*$ (with $E_*=\sqrt{n_0T_e}$) and the deepest ion density depletion $\delta n_i^{\max}/n_0$ inside a representative collapsing packet. Driven by ponderomotive self-focusing, the packet contracts and its field envelope rises to a peak of $\mathcal E_{\max}/E_*\simeq1.78$ at $\omega_p t=1090$. At this point, kinetic damping initiates collapse arrest and rapidly depletes the wave field, removing roughly $96\%$ of the local electric energy by $\omega_p t=1550$. Because of ion inertia, the density depression lags behind the field and continues to deepen to $\delta n_i^{\max}/n_0\simeq0.30$ at $\omega_p t=1235$ before relaxing. Meanwhile, the packet width shrinks from about $38\lambda_{\rm De}$ to a minimum of $19\lambda_{\rm De}$ near $\omega_p t=1280$ and then gradually broadens again.

\begin{figure}[!t]
 \centering
 \includegraphics[width=.95\columnwidth]{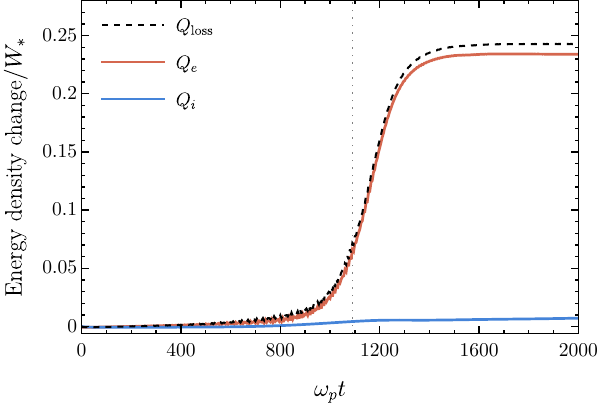}
 \caption{\label{fig:warpx-energy}Spatially averaged cumulative loss of wave energy density ($Q_{\rm loss}$) and corresponding gains in electron ($Q_e$) and ion ($Q_i$) thermal kinetic energy densities, in units of $W_*=n_0T_e/2$, from the 2D WarpX simulation. The vertical dotted line marks the onset of collapse arrest as in Fig.~\ref{fig:warpx}.}
\end{figure}

To follow where the dissipated wave energy goes, Fig.~\ref{fig:warpx-energy} tracks the spatially averaged energy budget relative to $\omega_p t=0$. Here $Q_{\rm loss}$ denotes the cumulative loss of wave energy density (combining the electric field energy and coherent electron oscillation energy), while $Q_e$ and $Q_i$ give the gains in random thermal kinetic energy density of electrons and ions, respectively. Between $\omega_p t=1090$ and $1550$, the wave energy density drops by $0.169W_*$, while $Q_e$ rises by $0.167W_*$ and $Q_i$ by $0.0016W_*$. Thus, roughly $99\%$ of the dissipated wave energy goes directly into electron heating, illustrating the collisionless collapse-dissipation channel modeled in the main text.

\section{Density-Scattering Model for Energy Injection and Transfer}\label{app:density-scattering}
\setcounter{equation}{0}\renewcommand{\theequation}{B\arabic{equation}}

\textbf{Mode energy budgets.}
To derive the steady-state injection and transfer coefficients used in the main text, we work directly with the driven Zakharov equations~\eqref{eq:physical-field} and \eqref{eq:physical-density}. We split the field envelope into its uniform mode $\bm{\mathcal E}_0=\avg{\bm{\mathcal E}}$ and finite-wavelength sidebands $\bm{\mathcal E}_{\rm s}=\bm{\mathcal E}-\bm{\mathcal E}_0$. Spatially averaging Eq.~\eqref{eq:physical-field} over a periodic domain with $\avg{\delta n}=0$ and vanishing uniform-mode damping, $\Gamma(0)=0$, yields
\begin{equation}
 \ii\partial_t\bm{\mathcal E}_0
 =\frac{\omega_p}{2}\bm{\mathcal E}_{\rm DP}
 +\frac{\omega_p}{2n_0}\avg{\delta n\,\bm{\mathcal E}_{\rm s}}.
 \label{eq:zero}
\end{equation}
We choose the phase of the linearly polarized drive so that $\bm{\mathcal E}_{\rm DP}=\mathcal E_{\rm DP}\eu$ is real, with unit polarization vector $\eu$ and amplitude $\mathcal E_{\rm DP}=\epsilon\sqrt{2\rho_{\rm DP}}$, and take the uniform mode along $\eu$, $\bm{\mathcal E}_0=\mathcal E_0\eu$. The uniform-mode energy density $W_0=\frac12|\mathcal E_0|^2$ then evolves according to $\partial_t W_0=\operatorname{Re}(\mathcal E_0^*\partial_t\mathcal E_0)=\mathcal P_{\rm in}-\mathcal P_{\rm tr}$, with
\begin{align}
 \mathcal P_{\rm in}&=-\frac{\omega_p}{2}\mathcal E_{\rm DP}\operatorname{Im}\mathcal E_0,\nonumber\\
 \mathcal P_{\rm tr}&=-\frac{\omega_p}{2n_0}\operatorname{Im}
   \bigl(\bm{\mathcal E}_0^*\cdot\avg{\delta n\,\bm{\mathcal E}_{\rm s}}\bigr).
 \label{eq:instantaneous-powers}
\end{align}
The density-coupling term $\mathcal P_{\rm L}(\delta n\,\bm{\mathcal E})$ transfers wave energy conservatively from the uniform mode to the sidebands at rate $\mathcal P_{\rm tr}$, so that the sideband energy density $W_{\rm s}=\frac12\avg{|\bm{\mathcal E}_{\rm s}|^2}$ obeys $\partial_t W_{\rm s}=\mathcal P_{\rm tr}-\mathcal P_{\rm loss}$, where $\mathcal P_{\rm loss}$ is the total sideband damping rate. In the quasi-steady state, where $\overline{\partial_t W_0}=\overline{\partial_t W_{\rm s}}=0$, late-time averaging directly recovers Eq.~\eqref{eq:balance}.

\smallskip
\textbf{Density-scattering response of the uniform mode.}
Although the uniform mode has no direct damping in Eq.~\eqref{eq:zero}, turbulent density perturbations continuously scatter it into finite-wavelength Langmuir sidebands. In Fourier space, a density mode $\delta n_{\bm k}$ scatters the uniform field $\mathcal E_0\eu$ into a longitudinal sideband $\bm{\mathcal E}_{\bm k}=\mathcal E_{\bm k}\hat{\bm k}$, where $\hat{\bm k}=\bm k/k$. Neglecting secondary sideband-to-sideband scattering in the $\bm k\ne0$ component of Eq.~\eqref{eq:physical-field}, each scattered mode evolves under thermal dispersion $\omega_k=\frac32\omega_p k^2\lambda_{\rm De}^2$ and damping $\Gamma(k)$ as
\begin{equation}
 \ii\partial_t\mathcal E_{\bm k}
 =[\omega_k-\ii\Gamma(k)]\mathcal E_{\bm k}
 +\frac{\omega_p(\hat{\bm k}\cdot\eu)}{2n_0}\delta n_{\bm k}\mathcal E_0,
\end{equation}
whose forced solution retains memory of the scattering over earlier time lags $s$:
\begin{align}
 \mathcal E_{\bm k}(t)&\simeq-\frac{\ii\omega_p(\hat{\bm k}\cdot\eu)}{2n_0}
 \int_0^\infty\dd s\,e^{-[\Gamma(k)+\ii\omega_k]s}\nonumber\\
 &\quad\times\delta n_{\bm k}(t-s)\mathcal E_0(t-s).
 \label{eq:scattered-sideband}
\end{align}
Here we neglect spontaneous background sidebands, whose random phases give no net contribution to $\overline{\avg{\delta n\,\bm{\mathcal E}_{\rm s}}}$ by symmetry.

These scattered sidebands interact again with the same density perturbations and scatter back onto the $k=0$ mode through $\avg{\delta n\,\bm{\mathcal E}_{\rm s}}$ in Eq.~\eqref{eq:zero}. Replacing $\mathcal E_0(t-s)$ by its steady mean $\overline{\mathcal E}_0$ over the sideband memory time, this back-reaction acts as a linear complex feedback, $\eu\cdot\frac{\omega_p}{2n_0}\overline{\avg{\delta n\,\bm{\mathcal E}_{\rm s}}}\simeq -\ii\zeta_D\overline{\mathcal E}_0$, with
\begin{align}
 \zeta_D&=\frac{\omega_p^2}{4n_0^2}\sum_{\bm k\ne0}(\hat{\bm k}\cdot\eu)^2
 \int_0^\infty\dd s\,e^{-[\Gamma(k)+\ii\omega_k]s}\nonumber\\
 &\quad\times\overline{\delta n_{\bm k}^*(t)\delta n_{\bm k}(t-s)}.
 \label{eq:zd-def}
\end{align}
In Eq.~\eqref{eq:zero}, $\operatorname{Re}\zeta_D$ acts as an effective scattering damping rate that drains energy from the uniform mode into the sidebands, while $\operatorname{Im}\zeta_D$ shifts the uniform-mode frequency away from exact resonance with the dark photon drive. Balancing the drive against this feedback gives the steady-state uniform field $\overline{\mathcal E}_0\simeq-\ii\omega_p\mathcal E_{\rm DP}/(2\zeta_D)$. Because $\operatorname{Im}\zeta_D\ne0$, $\overline{\mathcal E}_0$ is no longer purely out of phase with the drive, reducing power absorption by the phase factor $\operatorname{Re}\zeta_D/|\zeta_D|$. When uniform-mode fluctuations about the mean are small, so that $|\overline{\mathcal E}_0|\simeq\sqrt{2\overline W_0}$, Eq.~\eqref{eq:instantaneous-powers} yields
\begin{align}
 \overline{\mathcal P}_{\rm tr}&\simeq 2\overline W_0\operatorname{Re}\zeta_D,\nonumber\\
 \overline{\mathcal P}_{\rm in}&\simeq \frac{\omega_p}{2}\mathcal E_{\rm DP}\sqrt{2\overline W_0}\,\frac{\operatorname{Re}\zeta_D}{|\zeta_D|}.
 \label{eq:feedback-powers}
\end{align}

In Eq.~\eqref{eq:zd-def}, a scattered sideband remains coherent with the density perturbation that generated it only until wave damping, thermal dispersion, or the evolution of the density perturbation itself breaks their correlation. Modeling this finite density-mode lifetime as $\overline{\delta n_{\bm k}^*(t)\delta n_{\bm k}(t-s)}\simeq\overline{|\delta n_{\bm k}|^2}e^{-s/\tau_k}$, where $\tau_k$ is the integrated correlation time, gives
\begin{equation}
 \zeta_D\simeq\frac{\omega_p^2}{4n_0^2}\sum_{\bm k\ne0}
 \frac{(\hat{\bm k}\cdot\eu)^2\,\overline{|\delta n_{\bm k}|^2}}
 {\Gamma(k)+\tau_k^{-1}+\ii\omega_k}.
 \label{eq:zd-sum}
\end{equation}
Thus, even across long-wavelength modes where linear damping is negligible, $\Gamma(k)\ll\tau_k^{-1}$, the finite lifetime $\tau_k$ of evolving density depressions produces a nonzero $\operatorname{Re}\zeta_D$ and sustains irreversible energy transfer to the sidebands, while thermal dispersion $\omega_k$ generates $\operatorname{Im}\zeta_D$ and fixes the phase factor $\operatorname{Re}\zeta_D/|\zeta_D|<1$.

\smallskip
\textbf{Density fluctuations of collapsing wave packets.}
To evaluate Eq.~\eqref{eq:zd-sum}, we model the turbulent density field as a superposition of independently nucleated collapsing wave packets at random centers. Each packet contracts from an initial width $\ell_i$ to an arrest width $\ell_f$, with representative values $\ell_f=10\lambda_{\rm De}$ in 2D and $20\lambda_{\rm De}$ in 3D~\cite{RN}. These widths define the characteristic wavenumbers $k_i=1/\ell_i$ and $k_f=1/\ell_f$. At nucleation, the central wave energy density is $c_i\overline W_{\rm s}$, with the typical peak-to-mean ratio $c_i=5$~\cite{Rob}, and satisfies the collapse threshold $c_i\overline W_{\rm s}\ell_i^2=4\Theta_D W_*\lambda_{\rm De}^2$, where $\Theta_2=70$ and $\Theta_3=260$~\cite{RN}. Thus, stronger sideband turbulence forces wave packets to nucleate at smaller scales, $\ell_i\propto\overline W_{\rm s}^{-1/2}$.
For $k\ll k_i$, both the density power per mode and the cavity correlation time approach finite limits, so the feedback per logarithmic wavenumber interval decreases as $k^D$. For $k\gg k_f$, the Gaussian cavity profile suppresses the density spectrum. We therefore use $[k_i,k_f]$ as an effective integration band for an order-of-magnitude estimate of the feedback coefficient.

For isotropic turbulence in $D$ dimensions, averaging $(\hat{\bm k}\cdot\eu)^2$ over directions gives $1/D$. Adopting the strong-turbulence density spectrum $\overline{|\delta n_{\bm k}|^2}\propto k^{2-2D}$~\cite{Rob,RNdensity} and replacing the sum in Eq.~\eqref{eq:zd-sum} by an integral over $[k_i,k_f]$ normalized to the total relative density variance $V_n=\overline{\avg{(\delta n/n_0)^2}}=n_0^{-2}\sum_{\bm k\ne0}\overline{|\delta n_{\bm k}|^2}$ yields
\begin{equation}
 \zeta_D(\overline W_{\rm s})\simeq\frac{\omega_p^2 V_n}{4D}
 \frac{\displaystyle\int_{k_i}^{k_f}
       \frac{k^{1-D}\,\dd k}{\Gamma(k)+\tau_k^{-1}+\ii\omega_k}}
      {\displaystyle\int_{k_i}^{k_f}k^{1-D}\,\dd k},
 \label{eq:feedback-integral}
\end{equation}
whose dependence on $\overline W_{\rm s}$ enters through $V_n$, $k_i$, and $\tau_k$.

To determine $V_n$ and $\tau_k$, we describe the contraction of each packet by the subsonic collapse trajectory~\cite{RN,Rob}
\begin{equation}
 \ell^2(t_p)\simeq\ell_i^2-\omega_p\lambda_{\rm De}^2\,t_p,
 \qquad t_f\simeq\frac{\ell_i^2-\ell_f^2}{\omega_p\lambda_{\rm De}^2},
 \label{eq:trajectory}
\end{equation}
where $t_p\in[0,t_f]$ is the packet age and $t_f$ is its lifetime. We adopt a Gaussian wave-energy profile $W(\bm x,t_p)=c_i\overline W_{\rm s}(\ell_i/\ell)^2 e^{-x^2/\ell^2}$, whose peak rises as $\ell^{-2}$ to keep ponderomotive self-focusing balanced against thermal dispersion. In the quasistatic limit of Eq.~\eqref{eq:physical-density}, this traps the wave packet in a progressively deeper density cavity, $\delta n(\bm x,t_p)/n_0\simeq -W(\bm x,t_p)/(4\beta^2 W_*)$, where $\beta^2=1+3T_i/T_e$, which equals $4$ for our isothermal plasma with $T_i=T_e$.

In steady turbulence, active wave packets are uniformly distributed in age over $t_p\in[0,t_f]$. We take collapsing cores to carry all of the sideband energy in 2D and half of it in 3D, so that the packet number density is $N_P=\overline W_{\rm s}/\langle\int\dd^2 x\,W\rangle_{t_p}$ in 2D and $N_P=(\overline W_{\rm s}/2)/\langle\int\dd^3 x\,W\rangle_{t_p}$ in 3D. Because the packets are independently located, the total relative density variance is $V_n=N_P\langle\int\dd^D x\,(\delta n/n_0)^2\rangle_{t_p}$. Evaluating these Gaussian integrals and changing variables to $\ell$ via Eq.~\eqref{eq:trajectory}, we obtain in the wide-band limit $\ell_i\gg\ell_f$
\begin{equation}
 V_n\simeq\frac{c_i\overline W_{\rm s}^2}{16\beta^4 W_*^2}
 \begin{cases}
 \ln(\ell_i/\ell_f), & D=2,\\[5pt]
 \dfrac{3}{4\sqrt{2}}, & D=3.
 \end{cases}
 \label{eq:density-variance}
\end{equation}
Here $V_n\propto\overline W_{\rm s}^2$ because stronger sideband waves dig proportionally deeper cavities, $\delta n\propto\overline W_{\rm s}$. Moreover, the $\ln(\ell_i/\ell_f)$ factor in 2D reflects the logarithmic accumulation of density variance toward small scales, whereas in 3D the variance integral converges to a constant for $\ell_i\gg\ell_f$.

To determine the correlation time $\tau_k$, we evaluate the two-time correlation $\overline{\delta n_{\bm k}^*(t)\delta n_{\bm k}(t-s)}$ of the turbulent density modes. For a single collapsing Gaussian cavity centered at the origin, $\delta n(\bm x,t_p)\propto\ell(t_p)^{-2}e^{-x^2/\ell(t_p)^2}$, the $D$-dimensional spatial Fourier transform evolves as $h_k(t_p)=\ell(t_p)^{D-2}e^{-k^2\ell(t_p)^2/4}$ for $t_p\in[0,t_f]$ and vanishes otherwise, so a wave number $k$ is excited primarily as the cavity contracts through $\ell\sim 1/k$. Shifting a cavity to a random center $\bm x_a$ multiplies its Fourier mode by $e^{-\ii\bm k\cdot\bm x_a}$, so a superposition of independently nucleated cavities born at random times $t_a$ has Fourier amplitude
\begin{equation}
 \delta n_{\bm k}(t)\propto\sum_a e^{-\ii\bm k\cdot\bm x_a}\,h_k(t-t_a).
\end{equation}
Forming the two-time product $\delta n_{\bm k}^*(t)\delta n_{\bm k}(t-s)$ yields a double sum over cavity pairs with phase factors $e^{\ii\bm k\cdot(\bm x_a-\bm x_b)}$. Over many independent nucleation events, cross terms with $a\ne b$ carry random phases from uncorrelated cavity centers and cancel on average, leaving only the $a=b$ self-overlap terms. Averaging over stationary nucleation times then replaces the remaining sum by an integral over packet age $t_p=t-t_a$,
\begin{equation}
 \overline{\delta n_{\bm k}^*(t)\delta n_{\bm k}(t-s)}
 \propto\int_0^{t_f}h_k(t_p)\,h_k(t_p-s)\,\dd t_p.
\end{equation}
Evaluating the integrated correlation time defined by the exponential model in Eq.~\eqref{eq:zd-sum} then yields
\begin{align}
 \tau_k
 &=\frac{\int_0^\infty\overline{\delta n_{\bm k}^*(t)\delta n_{\bm k}(t-s)}\,\dd s}
        {\overline{|\delta n_{\bm k}|^2}}\nonumber\\
 &=\frac{\int_{-\infty}^\infty\dd s\int_0^{t_f}\dd t_p\,
         h_k(t_p)\,h_k(t_p-s)}
        {2\int_0^{t_f}h_k(t_p)^2\,\dd t_p}\nonumber\\
 &=\frac{\left(\int_0^{t_f}h_k\,\dd t_p\right)^2}
        {2\int_0^{t_f}h_k^2\,\dd t_p}.
 \label{eq:density-memory}
\end{align}
Physically, $\tau_k$ is the duration over which a collapsing cavity sweeps past the scale $\ell\sim 1/k$. From Eq.~\eqref{eq:trajectory}, subsonic contraction paced by thermal dispersion crosses this scale on the timescale $\tau_k\sim\ell^2/(\omega_p\lambda_{\rm De}^2)\sim(\omega_p k^2\lambda_{\rm De}^2)^{-1}\sim\omega_k^{-1}$, so each density mode stays coherent for roughly one dispersive period of the scattered sideband. Because the decorrelation rate $\tau_k^{-1}$ and the dispersive frequency shift $\omega_k$ are therefore comparable across the feedback band, $\operatorname{Re}\zeta_D$ and $|\operatorname{Im}\zeta_D|$ are of the same order, making the phase factor $\operatorname{Re}\zeta_D/|\zeta_D|$ an order-unity constant.

\smallskip
\textbf{Injection and transfer coefficients.}
With $V_n$ in Eq.~\eqref{eq:density-variance} and $\tau_k$ in Eq.~\eqref{eq:density-memory} determined, we can evaluate the feedback integral in Eq.~\eqref{eq:feedback-integral}. Because $\tau_k\sim\omega_k^{-1}\propto k^{-2}$, the integral is dominated by the long-wavelength end $k\sim k_i=1/\ell_i$, where linear damping is negligible, $\Gamma(k)\ll\tau_k^{-1}$. The feedback rate therefore scales as $\zeta_D\sim\omega_p^2 V_n\tau_{k_i}$. As $\overline W_{\rm s}$ increases, deeper cavities raise the scattering strength as $V_n\propto\overline W_{\rm s}^2$, while smaller nucleation scales $\ell_i\propto\overline W_{\rm s}^{-1/2}$ shorten the coherence time as $\tau_{k_i}\propto\ell_i^2\propto\overline W_{\rm s}^{-1}$. Their product yields a linear scaling,
\begin{equation}
 \zeta_D(\overline W_{\rm s})\simeq K_D\,\omega_p\frac{\overline W_{\rm s}}{W_*},
\end{equation}
where evaluating Eq.~\eqref{eq:feedback-integral} in the wide-band limit $\ell_i\gg\ell_f$ gives the dimensionless complex constant $K_2\simeq0.0183-0.0312\ii$ in 2D and $K_3\simeq0.0174-0.0237\ii$ in 3D. Substituting this linear feedback rate into Eq.~\eqref{eq:feedback-powers} directly recovers Eqs.~\eqref{eq:steady-injection} and \eqref{eq:steady-transfer} of the main text, with
\begin{equation}
 C_{\rm in}=\frac{\operatorname{Re}K_D}{|K_D|},
 \qquad C_{\rm tr}=2\operatorname{Re}K_D,
 \label{eq:cin-ctr}
\end{equation}
yielding $C_{\rm tr}\simeq0.0366$ in 2D and $0.0347$ in 3D, together with $C_{\rm in}\simeq0.505$ in 2D and $0.590$ in 3D.

\end{document}